# Effect of phase change on shock wave and n-dodecane droplet interaction with numerical investigation

Jiaxi Song, Tian Long, Shucheng Pan*

*School of Aeronautics, Northwestern Polytechnical University, Xi'an 710072, China*

**Abstract**

In a real propulsion system, shock-droplet interaction is often accompanied by phase change, which has a significant effect on the deformation and fragmentation of the droplet. In this paper, we study the effect of phase change on the n-dodecane droplet propulsion, deformation and fragmentation impacted by shock waves with high-resolution numerical simulations. First, we conduct a comparative study on shock waves and n-dodecane droplets interaction with and without phase change model. The impact of the shock wave changes the pressure and temperature distribution around the droplet, causing the droplet liquefaction on the windward side. With the influence of phase change, the Kelvin-Helmholtz instability (KHI) waves on the windward surface are enhanced, the development of KHI waves on the leeward surface of droplet is inhibited by vaporization. Furthermore, it is found that phase change suppresses both the flattening of the cylinder and shearing of the sheet at droplet equator. Next, we investigate the effect of Mach number on shock-droplet interaction with consideration of phase change. As the shock Mach number increases, the flattening and vaporization of droplets are suppressed, the KHI waves on the windward surface and the shear stripping of the sheet at the droplet equator are enhanced. The shear stripping of the liquid sheet plays a more dominant role in the deformation and breakup process than the flattening of the droplet under the SIE breakup mechanism in a higher Mach number.

**Keywords:** Phase change, Vaporization, Shock-droplet interaction, Compressible multiphase flow, Numerical simulation

1. **Introduction**

The study of shock-droplet interaction has a variety of engineering and geophysical applications, including raindrop damage during supersonic flight [1], sprays [2], shock wave lithotripsy [3] and secondary atomization of liquid jets in supersonic combustion systems [4]. Especially, the vaporization, deformation and fragmentation of fuel droplets driven by shock waves in high-temperature environments plays a significant role in supersonic combustion ramjet engines and liquid fuel rotating detonation systems [5] [6]. Compared with gaseous fuel, the atomization and vaporization process of liquid fuel can effectively reduce the time available for the air to mix with the vaporized fuel and improve the utilization of fuel [7]. Therefore, it is critical to deeply understand the interaction process between the shock wave and fuel droplets in high-temperature environments. In a real propulsion system, the interaction between shock waves and droplets must be accompanied by phase change, which is the focus of this paper.

The deformation and fragmentation of a single droplet driven by shock waves has been extensively studied in the past few decades [8] [9] [10] [11] [12] [13] [14] [15] [16] [17] [18] [19] [20] [21] [22]. Due to the limitations of experimental conditions and numerical models, most studies have not considered the effect of phase change. In the context of shock-droplet interaction without phase change, the Weber number is the most important parameter controlling droplets deformation and fragmentation. Traditionally, there are five breakup mechanisms with increasing Weber number established by Pilch and Erdman [12], which are vibrational, bag, bag-and-stamen, stripping, and catastrophic mechanisms. Theofanous et al. [14] reclassified the breakup mechanisms by using laser-induced fluorescence to visualize droplets, which are Rayleigh-Taylor piercing (RTP) for low Weber numbers and shear-induced entrainment (SIE) for high Weber numbers as the terminal breakup mechanism. Furthermore, they pointed out that the catastrophic breakup mechanism does not exist, which is a mirage of the shadowgraphs used to visualize waves [14].

Several recent studies have suggested that the effect of phase change can have a significant impact on droplets deformation and breakup mechanisms, mainly including cavitation and vaporization. In terms of cavitation, Sembian et al. [23] used a detailed experimental analysis found that a focusing of an expansion wave resulted by the transmitted wave reflecting at the downstream interface could create negative pressures sufficient for initiating cavitation, especially for large droplets at high Mach numbers. To further study the effect of cavitation on droplet breakup, Xiang and Wang [24] performed a numerical study on the interaction of planar shock waves and water columns embedded with a cavity at high Weber numbers. It was found that the momentum of the transverse jet increases when increasing the embedded-cavity radius in the water column under the same shock strength. Afterwards, Liang et al. [25] experimentally investigated the interaction of shock waves and water droplets embedded with a vapor cavity for the first time. They used the direct high-speed photography to capture the clear experimental images and found that the relative size and eccentricity of the cavity have a great influence on the movement and deformation of the hollow droplet. In addition, the transverse jet found in Xiang and Wang [24] emerges from the upstream interface and impacts on the downstream interface, eventually leading to the appears of a water jet. However, the above studies did not give the formation process of cavitation inside the droplet. Up to now, both the numerical simulations and experimental studies of cavitation bubble formation inside the droplet is difficult to implement.

In the context of vaporization, most studies have focused on the breakup of vaporizing droplets for incompressible flow at low Mach numbers. Haywood et al. [27] [28] used a nonorthogonal adaptive grid to predict the evaporation and deformation of n-heptane droplets in a high-temperature air environment. The predictions based on existing Nusselt and Sherwood number correlations show a good agreement with the numerical results. The quasi-steady drag correlation based on the instantaneous projected frontal area can also predict the aerodynamic drag for droplets, which is a function of Reynolds and Weber numbers. Nevertheless, their numerical simulations are unable to accurately represent the fragmentation of droplets because of the limited resolution. Strotos et al. [29] used the volume of fluid model coupled with a local evaporation model and adaptive grid refinement to study the effect of heating and evaporation for a volatile n-heptane droplet breakup in a high temperature gas environment. They concluded that the effect of heating has a small impact on droplet breakup except for low Weber numbers, due to its short duration. On the contrary, droplets deformation and breakup could enhance the heat transfer and evaporation of droplets. In terms of high-speed compressible flow, Goossens et al. [30] were one of the first who examined the evaporation of droplets induced by shock waves. They found that the droplet evaporation rate can be described as a process governed by heat conduction and vapor diffusion for weak shock waves and relatively small droplet sizes. However, their studies did not attention to the droplet deformation and breakup behaviors. In recent years research, Das and Udaykumar [31] developed a sharp-interface method to calculate the vaporization of droplets in high-speed flows. Using this method, the physics of the vaporization of aluminum droplets in shocked flows is numerically investigated [32], which concluded that the Sherwood number and the Nusselt number of the droplet increase monotonically with the Reynolds number. However, as the Mach number increases from 1.1 to 3.5, the Sherwood number and the Nusselt number exhibit a non-monotonic behavior due to the transition from subsonic to supersonic. Then, they expanded the method to study the interaction of shocked flows and burning aluminum droplets [33]. Several reactive cases of shock-droplet interaction are performed to study the effect of Mach number and Reynolds number for reacting aluminum droplets. The results showed that aluminum droplets have a transition from diffusion limited to kinetically limited combustion as the Mach number is increased and the droplet size or the Reynolds number is decreased, which have a significantly influence on the flame dynamics and vaporization rate of the reacting droplets. Furthermore, they developed a new model for the Sherwood number spanning diffusion and kinetically limited regimes of the reacting aluminum droplets to predict the energy released from the combustion of aluminized energetic materials in shocked flows.

Up to now, there is still a lack of corresponding experimental data on the interaction between shock waves and evaporating droplets in high-temperature environments. In a recent study, Redding and Khare [34] firstly performed a study of fundamental mechanisms for the deformation, fragmentation, and vaporization of n-dodecane droplets impacted by normal shock waves using volume of fluid coupled diffuse interface method. They used a thermal-mechanical-chemical equilibrium relaxation procedure to simulate the effect of phase change and compared the effect of vaporization on breakup physics with and without the vaporization model, the results showed that there are some differences from the non-vaporizing droplets, the vaporization could suppress the interfacial instabilities on droplets. In addition, they found that the rate of vaporization is a function of the shock strength, low Mach number shock waves lead to higher vaporization. However, due to the limitation of the numerical method, the study of Redding and Khare [34] don't have a clear interface evolution for droplets, a more comprehensive study understanding the effect of phase change, covering complete droplet morphology evolution and propulsion laws of interaction phenomenon are still lacking.

Based on the reasons mentioned above, the purpose of this paper is to investigate the effect of phase change on droplet morphology evolution and propulsion impacted by shock waves. A fully conservative sharp-interface method for compressible multiphase flows with phase change [35] [36] [37] [38] is used to solve the interaction of shock waves and vaporing droplets. The thermophysical properties of n-dodecane are similar to aviation kerosene, so we choose n-dodecane as the object for fuel droplets and shock waves interaction. It is noteworthy that a real-fluid equation of state based on Helmholtz-energy is used for n-dodecane in this paper to obtain a more accurate result physically. Moreover, the effects of Mach number on shock wave and n-dodecane droplet interaction with phase change are studied at the end of the paper. The rest of this paper is organized as follows. The problem description and governing equations are specified in Section 2. The numerical model used in this paper, a fully conservative sharp-interface method for compressible multiphase flows with phase change is introduced in Section 3. Next, we verify the accuracy and grid independence of the numerical method in Section 4. Subsequently, the numerical results and discussions are presented in Section 5, and the conclusions are summarized in

Section 6.

## 2. Physical model

2.1. Problem description

The deformation and breakup of droplets usually occurs under the impact of high-speed airflow, which is typically generated by a planar shock wave due to its simplicity, robustness, and repeatability [20]. The simulations are carried out in a square computational domain given by $8D_0 \times 8D_0$, which is shown in Figure 1. The current choice of computational domain size is sufficient to balance the influence of domain boundary conditions and computational costs. The initial droplet diameter of $D_0 = 4.8$ mm is in line with previous studies [16] [19] [20] [22], the initial places of droplet center and shock wave are also given in Figure 1. In order to avoid contamination of the numerical solution by the shock wave reflected by the wall [39] [40], non-reflective boundary conditions are imposed for all domain boundaries.

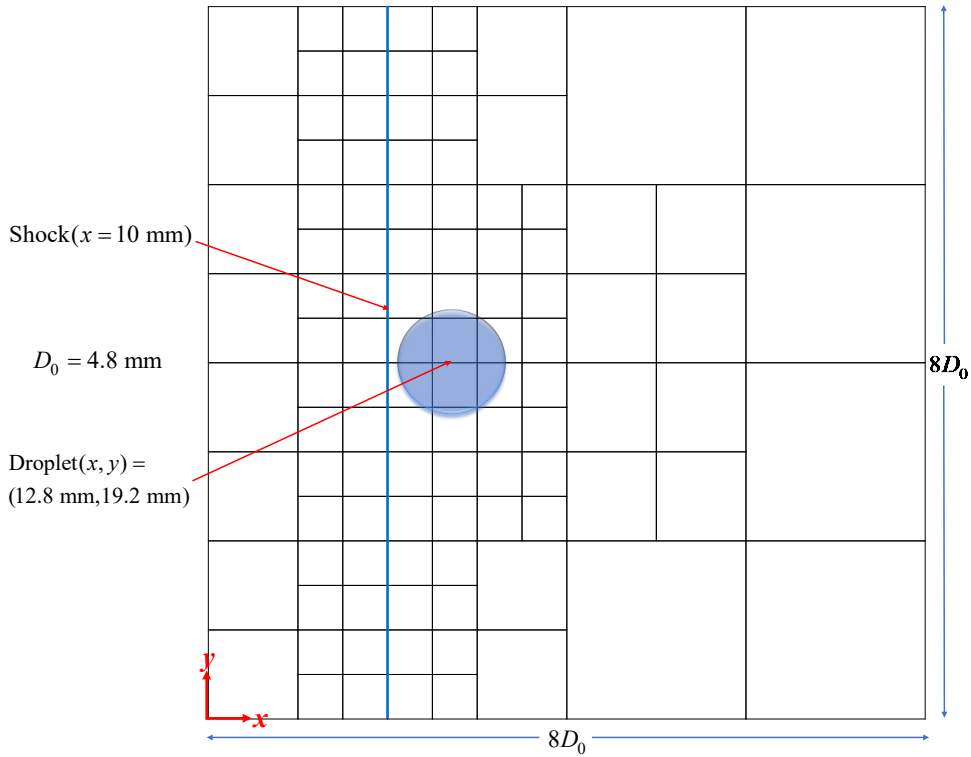

Figure 1. Schematic of the initial condition and computational domain for shock-droplet interaction. The droplet is sketched in blue circle, the shock in blue thick straight line, the squares represent different levels of blocks.

As the shock wave passes over the droplet surface, the high-speed airflow formed around the droplet causes the droplet to deform and breakup, and its behavior is dominated by inertial, viscous, and capillary forces. Inertial forces cause the deformation and breakup of droplets, viscous forces are just the opposite, and capillary forces are used to maintain the original spherical shape of the droplet. The interaction of the three forces described above on the droplet can be determined by the following dimensionless parameters. One of the most important parameters is the Weber number, which is defined as the ratio of inertial forces and capillary forces, that is

$$We = \frac{\rho_g u_g^2 D_0}{\sigma}, \qquad (1)$$

where $\rho_g$ and $u_g$ indicate the post-shock density and velocity of the gaseous phase. $D_0$ and $\sigma$ denote the initial droplet diameter and surface-tension coefficient respectively. Another equally important parameter is the Ohnesorge number, which is defined as the ratio of viscous forces and capillary forces, that is

$$Oh = \frac{\mu_l}{\sqrt{\rho_l D_0 \sigma}}, \qquad (2)$$

with the dynamic viscosity of the liquid droplet phase $\mu_l$ and the density of the liquid droplet phase $\rho_l$. The third is the Reynolds number, which is the ratio of the inertial forces to the viscous forces and defined as

$$\text{Re} = \frac{\rho_g u_g D_0}{\mu_g}, \qquad (3)$$

where $\mu_g$ indicates the dynamic viscosity of the gaseous phase. Furthermore, this paper also involves another dimensionless parameter, that is the shock-speed Mach number, defined as the ratio of the shock speed $u_s$ and the speed of sound $a$ in gaseous phase by

$$M_s = \frac{u_s}{a}. \qquad (4)$$

2.2. Governing equations

The conservation equations for two-phase viscous flows with phase change and surface tension term can be written as

$$\frac{\partial \mathbf{U}}{\partial t} + \nabla^T \cdot \mathbf{F} + \nabla^T \cdot \mathbf{F}_v = \mathbf{S}, \qquad (5)$$

where

$$\mathbf{U} = \begin{Bmatrix} \rho \\ \rho u \\ \rho v \\ \rho(e + V^2/2) \end{Bmatrix}, \qquad (6)$$

$$\mathbf{F} = \begin{Bmatrix} \rho u & \rho v \\ \rho u^2 + p & \rho uv \\ \rho uv & \rho v^2 + p \\ \rho(e+V^2/2)u + pu & \rho(e+V^2/2)v + pv \end{Bmatrix}, \qquad (7)$$

$$\mathbf{F}_v = -\begin{Bmatrix} 0 & 0 \\ \tau_{xx} & \tau_{xy} \\ \tau_{yx} & \tau_{yy} \\ k\frac{\partial T}{\partial x} + u\tau_{xx} + v\tau_{xy} & k\frac{\partial T}{\partial y} + u\tau_{yx} + v\tau_{yy} \end{Bmatrix}, \qquad (8)$$

denote the vector of flow state variables, the convective flux tensor and the viscous flux tensor, respectively. $\rho$, $u$, $v$, $e$, $p$ and $V$ are the density, the velocity component in the direction $x$ and $y$, internal energy, pressure and the modulus of velocity. $\tau_{xx}$, $\tau_{xy}$, $\tau_{yx}$ and $\tau_{yy}$ are the shear stress tensor, $k\frac{\partial T}{\partial x}$ and $k\frac{\partial T}{\partial y}$ are the heat flux component in the direction x and y. It is worth noting that the vector $\mathbf{S}$ denotes exchange terms between the liquid phase and gaseous phase including phase change, surface tension and viscous effects, for more details see the following Section 3.

### 2.3. Equations of state

The system of governing equations needs to be closed by the equations of state (EOS). In this paper, we consider three kinds of fluids, which are air, water and n-dodecane. Among that, the fluids of air and water are described by the stiffened-gas EOS [41] [42], which is given by

$$e(p,\rho) = \frac{p + \gamma\pi}{\rho(\gamma - 1)} + q,$$
$$T(p,\rho) = \frac{p + \pi}{C_v\rho(\gamma - 1)}, \qquad (9)$$

where $e$ and $T$ are the internal energy and temperature, which are functions of the pressure $p$ and density $\rho$. $\gamma$, $\pi$, $q$, $C_v$ are the adiabatic coefficient, the parameter accounting for the pre-compression of the fluid, the reference internal energy and the heat capacity at constant volume, respectively. Following Ref. [41] and [42], the parameters of the stiffened-gas EOS for air, vapor and liquid water are shown in Table 1. When the phase of air and water vapor are considered with $\pi = 0$, the stiffened-gas EOS degenerates to the ideal-gas EOS.

Table 1. The stiffened-gas EOS parameters for air, vapor and liquid water.

| Fluid types | $\gamma$ | $\pi$ (Pa) | $q$ (J/kg) | $C_v$ (J/kg/K) |
| --- | --- | --- | --- | --- |
| Air | 1.4 | 0 | 0 | $0.718 \times 10^3$ |
| Water vapor | 1.33 | 0 | 1.99×10⁶ | 1.399×10³ |
| Water liquid | 2.35 | 10⁹ | -1.167×10⁶ | 1.816×10⁶ |

For n-dodecane fluids, a real-fluid EOS based on Helmholtz-energy [43] is used to obtain more accurate thermodynamic variables physically. By using this real-fluid EOS, the prediction error in the thermodynamic parameters of n-dodecane is reduced less than 1% [44], the specific formulation can be referred to Ref. [43].

## 3. Numerical model

### 3.1. Conservative discretization

The governing equations above is discretized by a finite-volume approach on Cartesian square cells in the domain. As illustrated in Figure 2, the domain is divided into two subdomains by the zero level-set function $\phi(\mathbf{x}) = 0$, which are the liquid subdomain $\phi(\mathbf{x}) < 0$ and the gaseous subdomain $\phi(\mathbf{x}) > 0$, respectively.

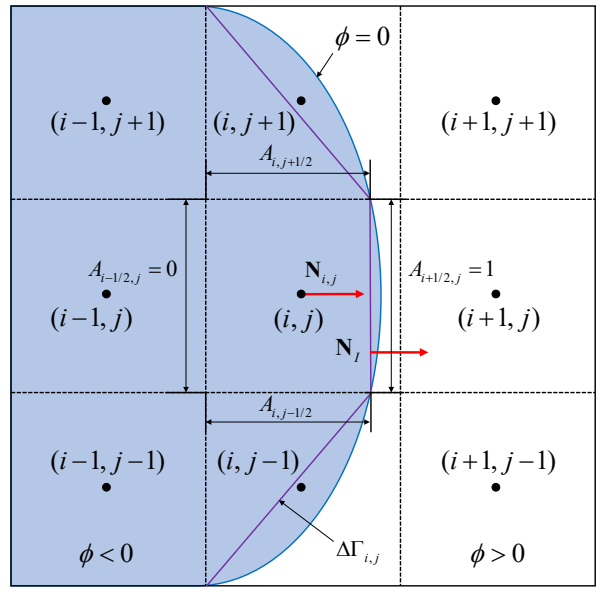

Figure 2. Two-dimensional schematic of the conservative discretization in a cut cell. The blue line denotes the exact interface segment, the purple line denotes the linearized approximation by the level-set function, the red arrow indicates the normal vector at the cell center.

Applying the Gauss's theorem and the first-order forward Euler method, the governing equations above can be written as

$$\alpha_{i,j}^{n+1}\mathbf{U}_{i,j}^{n+1} - \alpha_{i,j}^{n}\mathbf{U}_{i,j}^{n} = \frac{\Delta t}{\Delta x \Delta y}\mathbf{X}(\Delta\Gamma_{i,j}) +$$
$$\frac{\Delta t}{\Delta x}\left[A_{i-1/2,j}(\mathbf{F}_{i-1/2,j}+\mathbf{F}_{v,i-1/2,j}) - A_{i+1/2,j}(\mathbf{F}_{i+1/2,j}+\mathbf{F}_{v,i+1/2,j})\right] + \quad (10)$$
$$\frac{\Delta t}{\Delta y}\left[A_{i,j-1/2}(\mathbf{F}_{i,j-1/2}+\mathbf{F}_{v,i,j-1/2}) - A_{i,j+1/2}(\mathbf{F}_{i,j+1/2}+\mathbf{F}_{v,i,j+1/2})\right],$$

where $\alpha_{i,j}^{n}\mathbf{U}_{i,j}^{n}$ is defined the vector of conservative states at the cell center of cell $(i,j)$ and time $n$ for each material. $\alpha_{i,j}$ is the volume fraction of the corresponding phase, $\mathbf{U}_{i,j}$ is the vector of cell-averaged states, $\Delta t$ is the time step, $\Delta x$ and $\Delta y$ represent the grid spacing in the direction $x$ and $y$, respectively. The aperture $A$ denotes the cell face after segmentation by the interface $\Gamma$ at the current time step. The inviscid numerical flux $\mathbf{F}$ is obtained by the fifth-order Weighted Essentially Non-Oscillatory scheme [45] on characteristic fluxes, split by the global Lax-Friedrich scheme [46]. The viscous numerical flux $\mathbf{F}_v$ is interpolated using a fourth-order central finite-difference scheme. The term $\mathbf{X}(\Delta\Gamma_{i,j})$ denotes the momentum and energy exchange between liquid phase and gaseous phase in a cut cell with the effect of inviscid, viscous, surface tension and phase change, which is obtained by solving a two-phase Riemann problem along the normal direction. Furthermore, a second-order strong stability-preserving Runge-Kutta scheme [47] is used to perform the time marching. The maximum admissible timestep size is determined by

$$\Delta t = CFL \cdot \min\left(\frac{\Delta x}{\sum |u_i \pm c|_{\infty}}, \frac{3}{14}\frac{\rho \Delta x^2}{\mu}, \sqrt{\frac{\rho_l + \rho_g}{8\pi\sigma}\Delta x^3}\right), \quad (11)$$

where $c$ is the speed of sound and $CFL = 0.5$ is set for all the simulations in this paper.

3.2. Interface capturing

In this paper, the level-set method [48] is used to capture the two-phase interface during the shock wave and droplet interaction. The two-phase domain is represented by a level-set function $\phi(\mathbf{x})$. As the Figure 2 shows, the liquid phase subdomain is represented by $\phi(\mathbf{x}) < 0$, the gaseous phase subdomain is represented by $\phi(\mathbf{x}) > 0$, the two-phase interface is represented by $\phi(\mathbf{x}) = 0$, $|\phi(\mathbf{x})|$ represents the normal signed distance of the cell center $\mathbf{x}$ to the two-phase interface. The level-set function $\phi(\mathbf{x})$ is evolved in time with the advection equation

$$\frac{\partial \phi}{\partial t} + u_\phi \mathbf{n}_\Gamma \cdot \nabla \phi = 0, \quad (12)$$

where $u_\phi$ denotes the level-set advection velocity, which is determined from a two-material Riemann

problem [35]. The interface normal $\mathbf{n}_\Gamma$ and interface curvature $\kappa$ can be computed by

$$\mathbf{n}_\Gamma = \frac{\nabla \phi}{|\nabla \phi|}, \quad \kappa = \nabla \cdot \frac{\nabla \phi}{|\nabla \phi|}. \tag{13}$$

After the advection step, the level-set advection equation needs to be reinitialized to maintain the signed distance property $|\nabla \phi| = 1$ with the re-initialization equation [49]

$$\frac{\partial \phi}{\partial \tau} + sign(\phi_0)(|\nabla \phi| - 1) = 0. \tag{14}$$

The cell-face fluxes near the interface are obtained by the ghost fluid method [50] to assure the sharp interface property of the method.

3.3. Interface interactions

In order to improve numerical stability while guaranteeing strict conservation and sharp interfacial properties of each fluid, the interaction term $\mathbf{X}(\Delta \Gamma_{i,j})$ is obtained by solving a two-material Riemann problem with phase change [38], which can be written as four terms

$$\mathbf{X}(\Delta \Gamma_{i,j}) = \mathbf{X}^v(\Delta \Gamma_{i,j}) + \mathbf{X}^c(\Delta \Gamma_{i,j}) + \mathbf{X}^s(\Delta \Gamma_{i,j}) + \mathbf{X}^p(\Delta \Gamma_{i,j}), \tag{15}$$

where the right terms denote viscous, inertial, surface-tension force and effects of phase change, respectively. The viscous flux across the two-phase interface $\Delta \Gamma_{i,j}$ is

$$\mathbf{X}^v(\Delta \Gamma_{i,j}) = (0, \tau \Delta \Gamma_{i,j} \mathbf{n}_\Gamma, \tau \Delta \Gamma_{i,j} \mathbf{n}_\Gamma \cdot \mathbf{u})^T, \tag{16}$$

with the interface viscous stress tensor

$$\boldsymbol{\tau} = \mu(-\tfrac{2}{3} \nabla \cdot \mathbf{u} \mathbf{I} + (\nabla \mathbf{u} + \nabla \mathbf{u}^T)). \tag{17}$$

The combination of the inertial term and the surface tension term can be written as

$$\mathbf{X}^{c,m}(\Delta \Gamma_{i,j}) + \mathbf{X}^{s,m}(\Delta \Gamma_{i,j}) = (0, \Delta \Gamma_{i,j} p_{\Gamma,m} \mathbf{n}_\Gamma, \Delta \Gamma_{i,j} p_{\Gamma,m} \mathbf{n}_\Gamma \cdot \mathbf{u}_\Gamma)^T, \tag{18}$$

where the subscript $m$ stands for liquid phase or gaseous phase. The pressure jump caused by surface tension and mechanical equilibrium is

$$\Delta p = p_{\Gamma,1} - p_{\Gamma,2} = \sigma \kappa, \tag{19}$$

where $\sigma$ is the surface tension coefficient. As in Long et al. [38], the phase change term $\mathbf{X}^p(\Delta \Gamma_{i,j})$ can be obtained by

$$\mathbf{X}^{p,m}(\Delta \Gamma_{i,j}) = \pm(j \Delta \Gamma_{i,j}, j \Delta \Gamma_{i,j} \mathbf{n}_\Gamma \cdot \mathbf{u}_\Gamma, j \Delta \Gamma_{i,j}(e_\Gamma + \tfrac{1}{2}|\mathbf{u}_\Gamma|^2)), \tag{20}$$

where $e_\Gamma$ is the internal energy of the phase interface, + and – are applied for the gaseous phase and

liquid phase, respectively. The term $\Delta \Gamma_{i,j}$ can be approximated by

$$\Delta \Gamma_{i,j} = \sqrt{(A_{i+1/2,j} - A_{i-1/2,j})^2 + (A_{i,j+1/2} - A_{i,j-1/2})^2}, \tag{21}$$

Here, we need an additional phase change model to evaluate the mass flux $j$. In this paper, we employ the Hertz-Knudsen relation [51]

$$j = \frac{1}{2\pi R_g}(\lambda_e \frac{p_{sat} \cdot T_l}{\sqrt{T_l}} - \lambda_c \frac{p_g}{\sqrt{T_g}}), \tag{22}$$

where $R_g$ denotes the specific gas constant, which is set to $461.52 \, \text{J}/(\text{kg} \cdot \text{K})$ for water and $48.81 \, \text{J}/(\text{kg} \cdot \text{K})$ for n-dodecane. $\lambda_e$ and $\lambda_c$ are the evaporation and condensation coefficients, which are set to 1.0 and 0.6 for all the shock-droplet interaction simulations in this paper, respectively. $T_l$ and $T_g$ are the temperature of the phase interface for liquid and gas, $p_g$ the pressure of the phase interface for gaseous phase, $p_{sat}$ the saturation pressure. In our simulations, the saturation pressure of water is calculated according to Ref. [51], the saturation pressure of n-dodecane can be obtained through the opensource library CoolProp [52].

3.4. Wavelet-based multi-resolution method

The numerical simulation of the interaction between shock waves and droplets necessitates the utilization of a high-resolution mesh to accurately capture the evolution of the droplet interface. To enhance the computational efficiency, we adopt a block-structured adaptive multi-resolution mesh refinement technique [36] [53]. As illustrated in Figure 1, the domain is divided into a number of square blocks, the blocks near the droplet interface and strong variations in the fluid field such as the shock wave will undergo refinement. It is worth noting that all the blocks have a fixed number of internal cells, the adaptive refinement of grid is achieved by refining the blocks on different levels. In this paper, we set the number of initial blocks to 1 and internal cells per block to 16 for all the simulations. The effective resolution is determined by the maximum level $L_{\max}$, the maximum number of cells $N_{\max}$ can be calculated by

$$N_{\max} = 2^{L_{\max}} \cdot 16, \tag{23}$$

more details on adaptive multi-resolution method can be found in Ref. [36].

3.5. Interface scale separation

The high-resolution numerical simulations of shock-droplet interaction especially for droplet breakup usually produces many non-resolved interface segments, such as isolated small droplets and thin filaments. The linear approximation of these non-resolved interface segments generated during the interface evolution may lead to numerical fluctuations or even numerical instabilities simulations. In this paper, an interface-scale-separation model based on the constrained stimulus-response procedure developed by Han et al. [54] and Luo et al. [55] is used to separate the resolvable and non-resolvable interface scales automatically.

**4. Method validation**

4.1. Shock-droplet interaction without phase change

The method validation and grid sensitivity analyses are conducted by simulating the case of shock wave and water droplet interaction without phase change from Kaiser et al. [16]. The simulation domain and setup are shown in Figure 1, the Mach number of the shock wave is 1.47. The material parameters of liquid water and air can be found in Table 1. The initial conditions for water droplet, pre-shock air, and the post-shock air condition from the Rankine-Hugoniot relation [56] are shown in Table 2.

Table 2. The initial conditions for water droplet, pre-shock air, and post-shock air.

| Condition types | $\rho$(kg/m$^3$) | $p$(Pa) | $u$(m/s) | $\mu$(Pa·s) | $\sigma$ (N/m) |
|---|---|---|---|---|---|
| Water droplet | 1000.0 | 1.0×10$^5$ | 0.0 | 1.0×10$^{-3}$ | 0.073 |
| Pre-shock air | 1.2 | 1.0×10$^5$ | 0.0 | 1.8×10$^{-5}$ | / |
| Post-shock air | 2.18 | 2.35×10$^5$ | 225.9 | / | / |

The grid sensitivity analyses are performed by four different maximum levels of $L_{\max}$ = 5, 6, 7, 8, respectively, which relate to the effective resolutions of 512×512, 1024×1024, 2048×2048, and 4096×4096, the cells per initial droplet diameter of 64, 128, 256, and 512. Figure 3 shows the droplet interface evolution of different resolutions and previous results. As the resolution increases from 64 to 512 cells per initial droplet diameter, more details like the liquid sheets and thin filaments can be resolved, the overall qualitative features of the breakup mechanism like flattening and stripping are similar. It is obvious that a minimum of 256 cells per initial droplet diameter are required to resolve the KHI waves formed on the surface of the droplet. Compare with the results of Kaiser et al. [16], our results have a better agreement with the corresponding experiment results especially for later stage ($t^* = 0.76$). For the resolution of 512 cells, we could capture the details of interface instabilities in early stage ($t^* = 0.14$) which is hard to observe in the experiments. It is worth noting that the time in this paper is non-

dimensionalized by a scaling of the literature [10], which is given by

$$t^* = t\frac{u_g}{D_0}\sqrt{\frac{\rho_g}{\rho_l}}, \quad (24)$$

where $\rho_l$ and $\rho_g$ are the density of droplet and gaseous phase behind the shock wave, respectively. $u_g$ denotes the velocity of post-shock gaseous phase, $D_0$ denotes the initial diameter of the droplet.

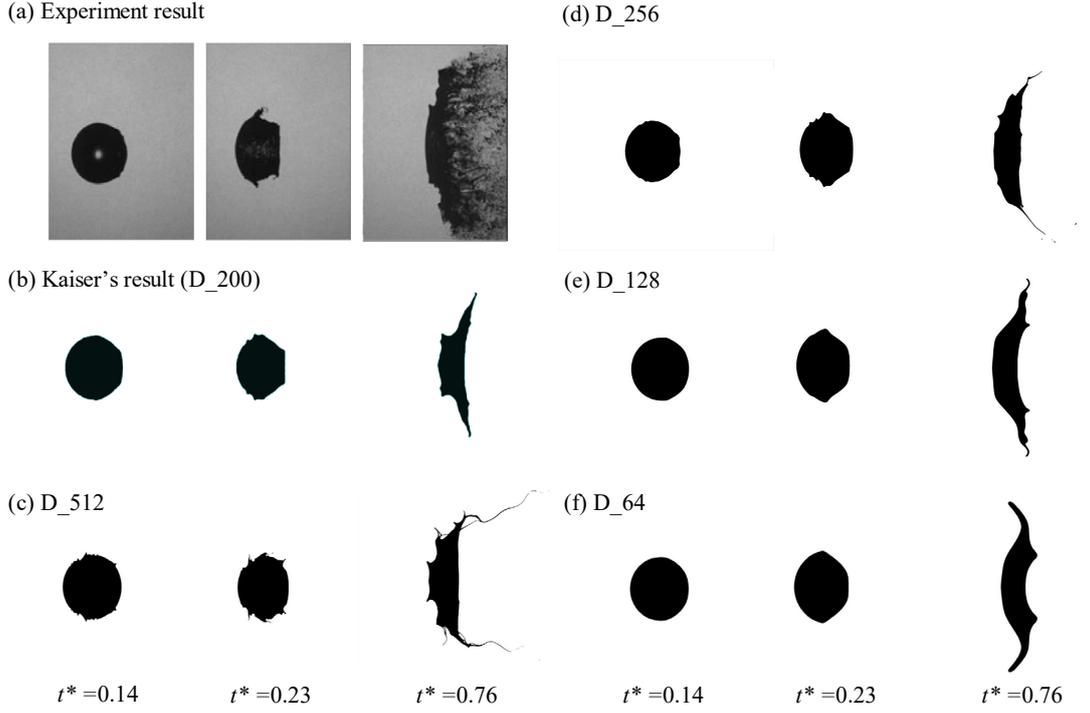

(a) Experiment result  (d) D_256
(b) Kaiser's result (D_200)  (e) D_128
(c) D_512  (f) D_64

$t^* =0.14$   $t^* =0.23$   $t^* =0.76$   $t^* =0.14$   $t^* =0.23$   $t^* =0.76$

Figure 3. Comparison of our simulation results for various resolutions, previous simulation results and the experimental results of water droplet interface evolution. (a) The experimental results are reprinted from Theofanous et al. [57]. (b) The simulation results from Kaiser et al. [16] with resolution of 200 cells per initial droplet diameter. (c)-(f) Our results of water droplet interface evolution for resolutions of 512, 256, 128, and 64 cells per initial droplet diameter, respectively.

The quantitative comparison of the dimensionless center-of-mass drift evolution is shown in Figure 4, the dimensionless center-of-mass location can be calculated using

$$x_c = \frac{\int_\Omega x\alpha_l\rho_l dV}{\int_\Omega \alpha_l\rho_l dV}, \quad x^* = \frac{\Delta x}{D_0} = \frac{x_c - x_0}{D_0}, \quad (25)$$

where $\alpha_l\rho_l$ denotes the liquid partial density, $x_0$ is initial location of the droplet. Note that there is a noteworthy deviation of the dimensionless center-of-mass drift between 64 cells and 1024 cells, more small droplets can be resolved with the increase of the cells per initial droplet diameter, which leads to a rearward drift of the center-of-mass. However, the evolution of center-of-mass drift between 512 cells and 1024 cells are almost overlap, which indicates the propulsion of the droplet are nearly independent of the grid size when 512 cells are arranged along the initial droplet diameter. In addition, our results also have a good agreement with Kaiser et al. [16] and Meng & Colonius [22].

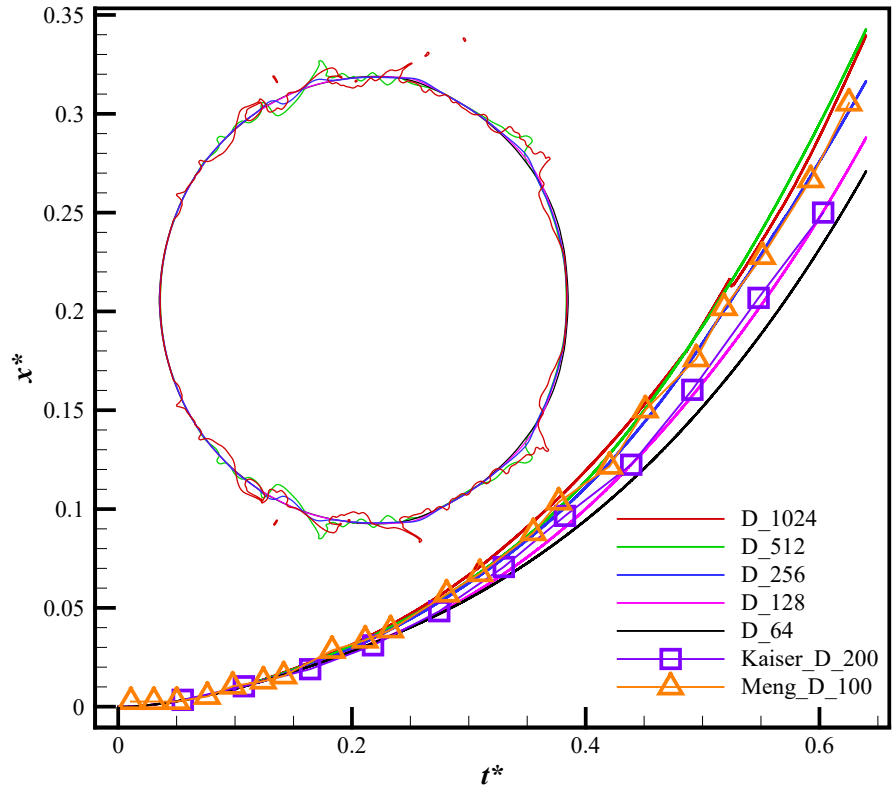

Figure 4. Comparison of our simulation results for various resolutions and previous simulation results of Kaiser et al. [16] and Meng & Colonius [22] for the dimensionless center-of-mass drift evolution.

In addition, Figure 5 shows the interface evolution of n-dodecane droplet under the 1.47 Mach number shock wave with consideration of phase change for various resolutions. Different form Figure 3, the resolution of 128 cells can resolve the effect of KHI waves which are enhanced by phase change. However, 512 cells will see a finer droplet interface. Combined with the analysis of the results in Figure 3, the resolution of 512 cells is used in the following simulations of n-dodecane droplet cases to resolve a more detailed interface evolution structure.

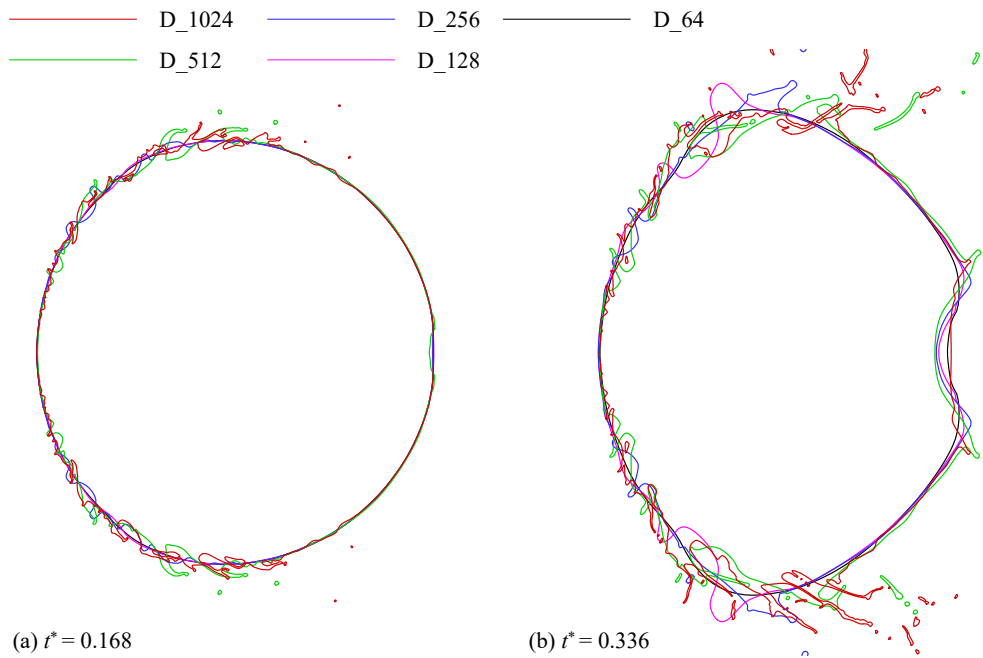

(a) $t^* = 0.168$  (b) $t^* = 0.336$

Figure 5. Comparison of the n-dodecane droplet morphology evolutions with phase change for resolutions of 1024, 512, 256, 128, and 64 cells per initial droplet diameter, respectively.

In the resolution of 512 cells, Figure 6 shows a comparison of experimental visualizations and numerical schlieren images, which indicates that the methods used in this study can precisely capture the characteristics of wave structures in the early stage of droplet breakup. The numerical simulation results are not only in good agreement with the experimental results in terms of the time evolution and flow field structure, but also can capture the finer flow field structure that cannot be observed experimentally.

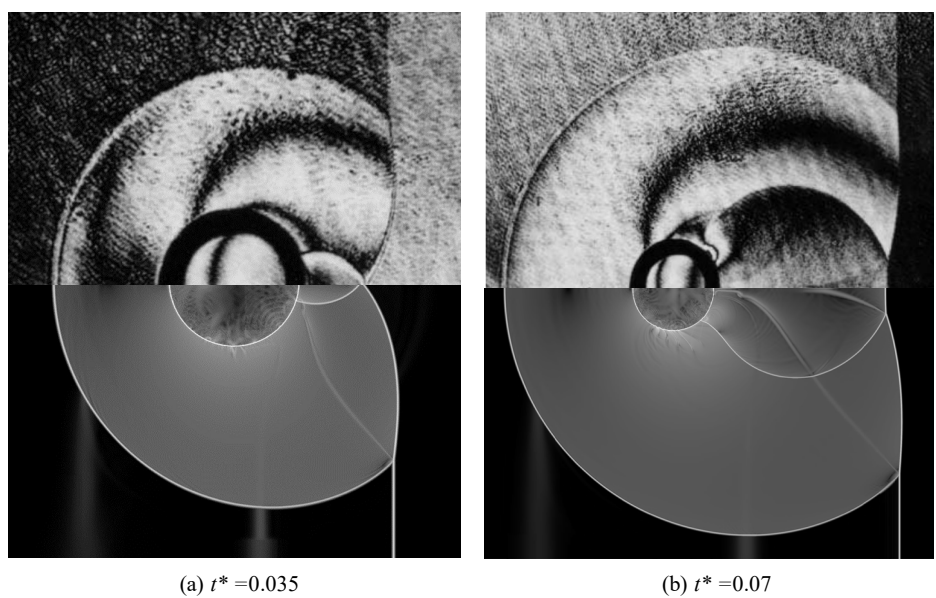

(a) $t^* = 0.035$  (b) $t^* = 0.07$

Figure 6. Comparison of the experimental visualizations [58] (upper half) and the numerical schlieren images

(lower half) with resolution of 512 cells per initial droplet diameter.

4.2. Droplet vaporization in a quiescent vapor environment

To verify the accuracy of the phase change model, we simulated the vaporization behavior of a single droplet in a high-temperature quiescent vapor environment. The computational domain is the same as in Figure 1, a singe water droplet with diameter of $D_0 = 4.8$ mm is placed at the center of the domain. The initial temperature of droplet and water vapor are 373 K and 380 K, respectively. Due to the high temperature of the vapor surrounding the droplet exceeding the boiling point, the droplet will vaporize. With the assumption of constant thermophysical properties, the time evolution of the droplet diameter follows the classical d² law [59], which can be written as

$$d^2(t) = d_0^2 - Kt, \qquad (26)$$

where $d_0$ denotes the initial diameter of the droplet, the constant $K$ is calculated by the fluid properties, more details can be found in Ref. [59].

Figure 7 shows that the numerical results are in good agreement with the analytical solution of d² law within 500 microseconds, which demonstrates the good accuracy of the phase change model on the time scale of the shock wave and droplet interaction.

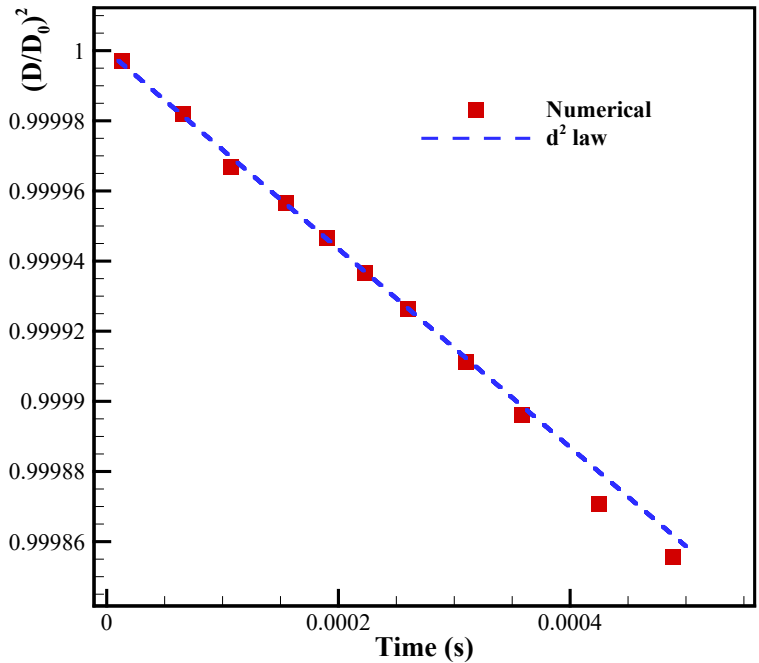

Figure 7. Comparison of the numerical result and analytical solution of d² law.

## 5. Results

5.1. Effect of phase change on shock-droplet interaction

To study the shock wave and n-dodecane droplet interaction with the influence of phase change, we simulated the deformation and fragmentation behavior of a n-dodecane droplet under the shock wave impact with and without consideration of phase change model respectively. The computational domain is shown in Figure 1, a liquid n-dodecane droplet impacted by a shock wave of 1.47 Mach in n-dodecane vapor environment. Table 3 lists the initial conditions of n-dodecane droplet and vapor for pre-shock and post-shock.

Table 3. The initial conditions for n-dodecane droplet, pre-shock vapor, and post-shock vapor.

| Condition types | $\rho(\mathrm{kg/m^3})$ | $p(\mathrm{Pa})$ | $u(\mathrm{m/s})$ | $T(\mathrm{K})$ | $\mu(\mathrm{Pa \cdot s})$ | $\sigma(\mathrm{N/m})$ |
|---|---|---|---|---|---|---|
| n-Dodecane droplet | 593.5 | $1.0\times10^5$ | 0.0 | 490.0 | $1.96\times10^{-4}$ | 0.009 |
| Pre-shock n-dodecane vapor | 4.38 | $1.0\times10^5$ | 0.0 | 500.0 | $7.59\times10^{-6}$ | / |
| Post-shock n-dodecane vapor | 10.15 | $2.35\times10^5$ | 123.45 | 510.49 | / | / |

When an incident shock wave propagates through a droplet, several different waves like reflected wave, transmitted wave, reflected expansion wave and Mach stem form, which have a significant impact on the flow field conditions around the droplet. Figure 8 shows the numerical schlieren images at the early stage of shock wave and n-dodecane droplet interaction with and without phase change, respectively.

The numerical schlieren images are calculated by $\log(|\nabla\rho|+1)$ to modify the scale of density gradient to accentuate the details of the wave structure [60]. Regardless of whether the phase change is considered, the wave system structure evolution of the shock wave impacting the droplet has a certain similarity. First, due to the nonlinear-acoustic mechanisms between the n-dodecane liquid droplet and the vapor interface, the transmitted shock wave and reflected shock wave are formd after the incident shock and droplet interaction [56]. It is worth noting that the reflected wave due to acoustic impedance mismatch is not always a shock wave, but the transmitted wave is always a shock wave [23]. Since only a fraction of the incident shock wave can penetrate the droplet surface, the intensity of the transmitted shock wave is much smaller than the reflected shock wave. Compared to the case without phase change, the wave system structure under phase change is much more complicated. On the one hand, the evaporation on the droplet surface makes the flow parameters jump in velocity and pressure at the interface, resulting in an increase in pressure and velocity on the vapor side of the droplet surface, forming an additional outwardly propagating evaporation shock wave [61]. On the other hand, due to the large temperature gradient at the interface between liquid droplet and vapor, the sudden increase in temperature of the vapor surrounding the droplet creates the thermally-induced shock wave [31], which can be seen in Figure 8(b). As the incident shock wave continues to move downstream of the droplet, the reflection process of the incident shock wave on the droplet surface will undergo a transition from regular reflection to Mach reflection, and the incident shock wave and reflected shock wave intersect at the triple point [23]. With the consideration of phase change, the interaction of Mach stem and thermally-induced shock wave results the generation of vortex structures on the droplet surface, we can see this in Figure 8(b) at time $t^* = 0.0619$. After this, the Mach-Mach collision occurs at the rear stagnation point, which initiates a secondary wave system. It is obvious that the recirculation zone with phase change has a more complicated vortex structure, which is formed by the interaction of Mach wave and thermally-induced shock wave.

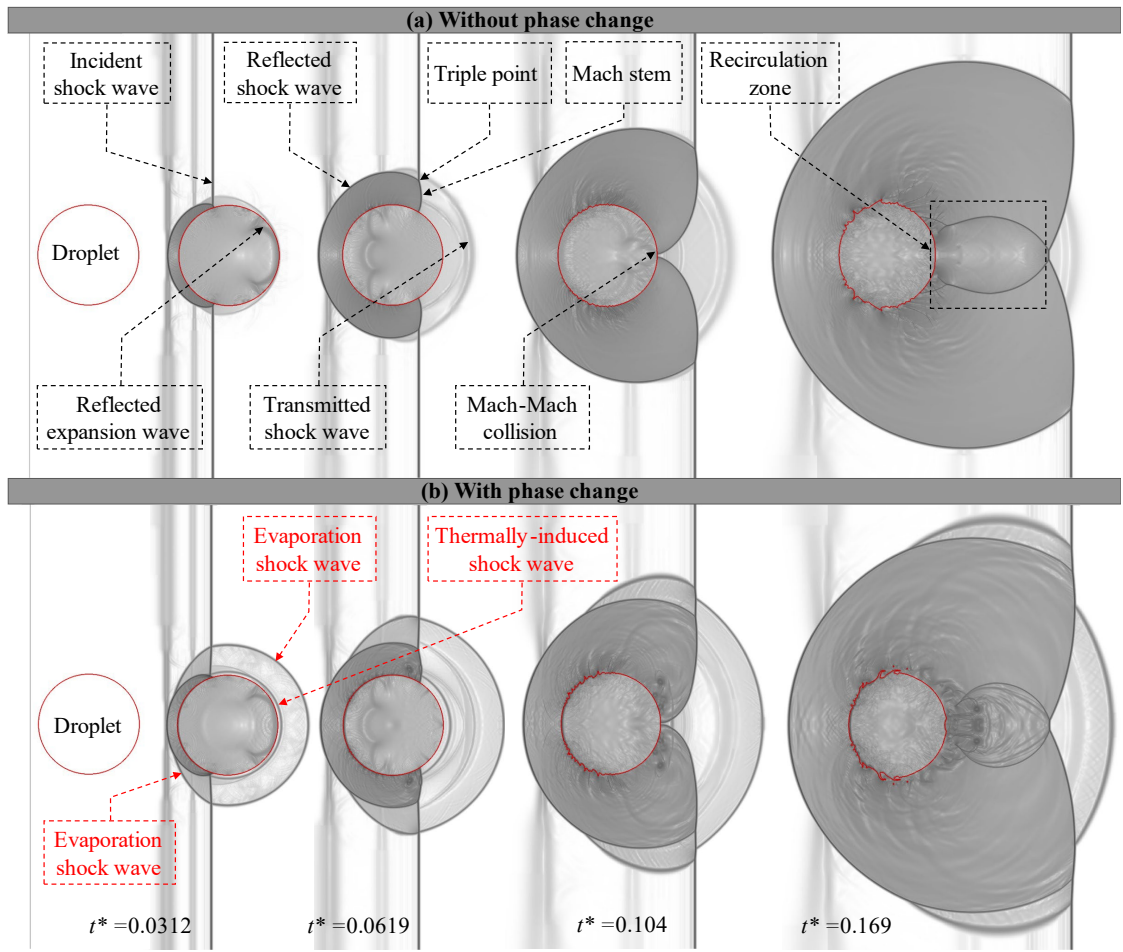

Figure 8. Comparison of numerical schlieren images for the logarithmic density gradient with and without phase change.

The differences induced by phase change are not only reflected in wave structures, but also in the evolution of the other physical quantities of the flow field. Figure 9 shows the temporal evolution of the temperature, pressure, and vorticity fields in two cases, where the blue lines illustrate the droplet interface. The introduction of phase transition changes the distribution of the temperature field on the leeward side of the droplet, a low temperature area close to the leeward side is formed due to vaporization. It can be seen from Figure 9(a) that the low-temperature region only formed on the leeward side, which indicates the vaporization occurs mainly on the leeward side of the droplet. Among them, Figure 10 clearly shows the mass flux distribution along the droplet interface impacted by the shock wave with phase change, from left to right are nine equal-time snapshots at $t^* = 0.1 \sim 0.9$. When the mass flux is positive, it means that the droplet surface is vaporizing, and a negative value of mass flux indicates that the droplet is liquefying. It is seen that the vaporization occurs mainly on the leeward side of the droplet, the closer to the equator, the stronger the vaporization effect on the surface of the droplet. For the windward side of the droplet, the distribution of phase change on droplet surface is transformed under the impact of shock wave, more liquefaction occurs on the windward side, especially in the early stages. Combined with Figure 9(b), there is a high pressure and temperature region created by the reflected shock wave on the windward side of the droplet. A higher pressure means the liquid n-dodecane molecules require a much higher kinetic energy to overcome the vapor pressure [34]. Thus, droplet molecules on the windward side may be more difficult to vaporize because they are exposed to a higher relative pressure than the leeward side. In this case, due to the initial n-dodecane vapor temperature is not high, the pressure in this region is greater than the saturated vapor pressure at the corresponding temperature, which causes n-dodecane vapor undergoes liquefies on the windward side of the droplet.

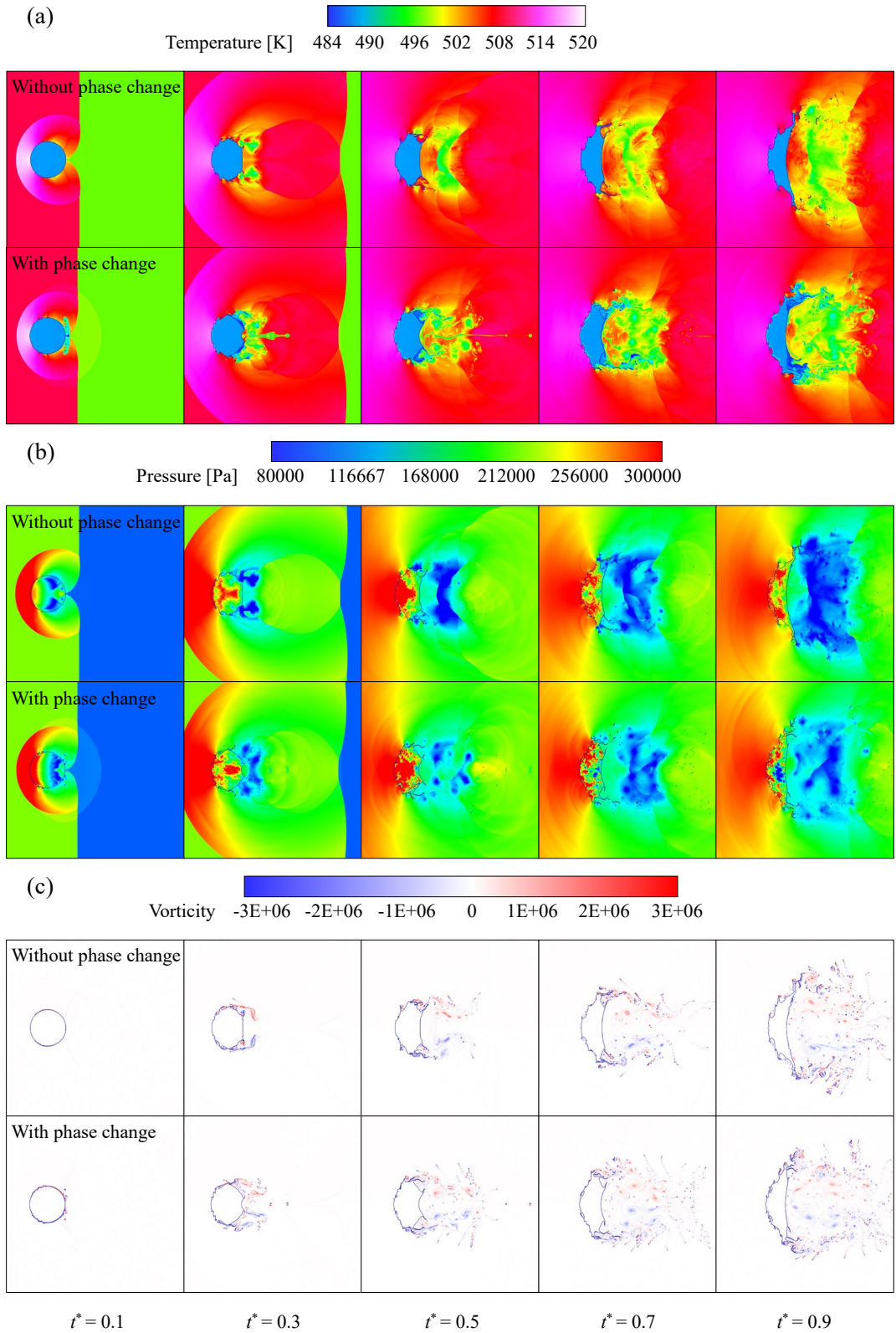

Figure 9. Comparison of the airflow temperature, pressure, and vorticity with and without phase change.

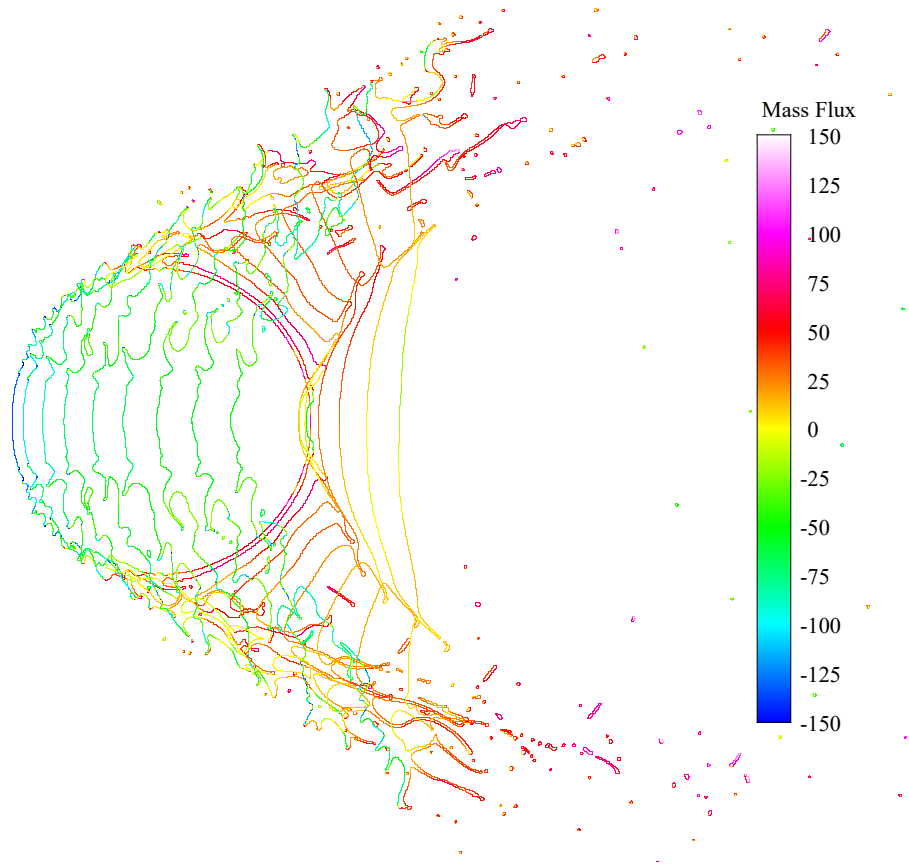

Figure 10. Mass flux distribution along the deformed droplet interface with phase change, from left to right are nine equal-time snapshots at $t^* = 0.1 \sim 0.9$.

Furthermore, due to the influence of phase change, there is a certain degree of enhancement in the vorticity of liquid droplets on the leeward side, especially in the early stages of droplet deformation, as shown in Figure 9(c). The additional vortices are generated as the interaction of Mach stem and thermally-induced shock wave, which can be seen in Figure 9(c) at time $t^* = 0.1$. The quantitative statistical study of vorticity is shown in Figure 11, where we compare the effect of phase change on the circulation along the total domain and droplet interface, respectively. Circulation is defined as the line intergral of velocity along a closed path $P$

$$\Gamma = \int_P u \cdot dl, \qquad (27)$$

which can be transformed into the surface integral $S$ over the vorticity by Stokes theorem

$$\Gamma = \int_S \omega \cdot dS. \qquad (28)$$

Thus, circulation magnitude is a macroscopic measure of the fluid rotation. As is shown in Figure 11 (a) and (b), compared to the case without phase change, the circulation for both total domain and droplet interface are increased, which means that more vorties are generated with the consideration of phase change. It is worth noting that the interface circulation increases significantly at the moment of 0.1, which corresponds to the moment of the Mach-Mach collision occurs, as shown in Figure 8.

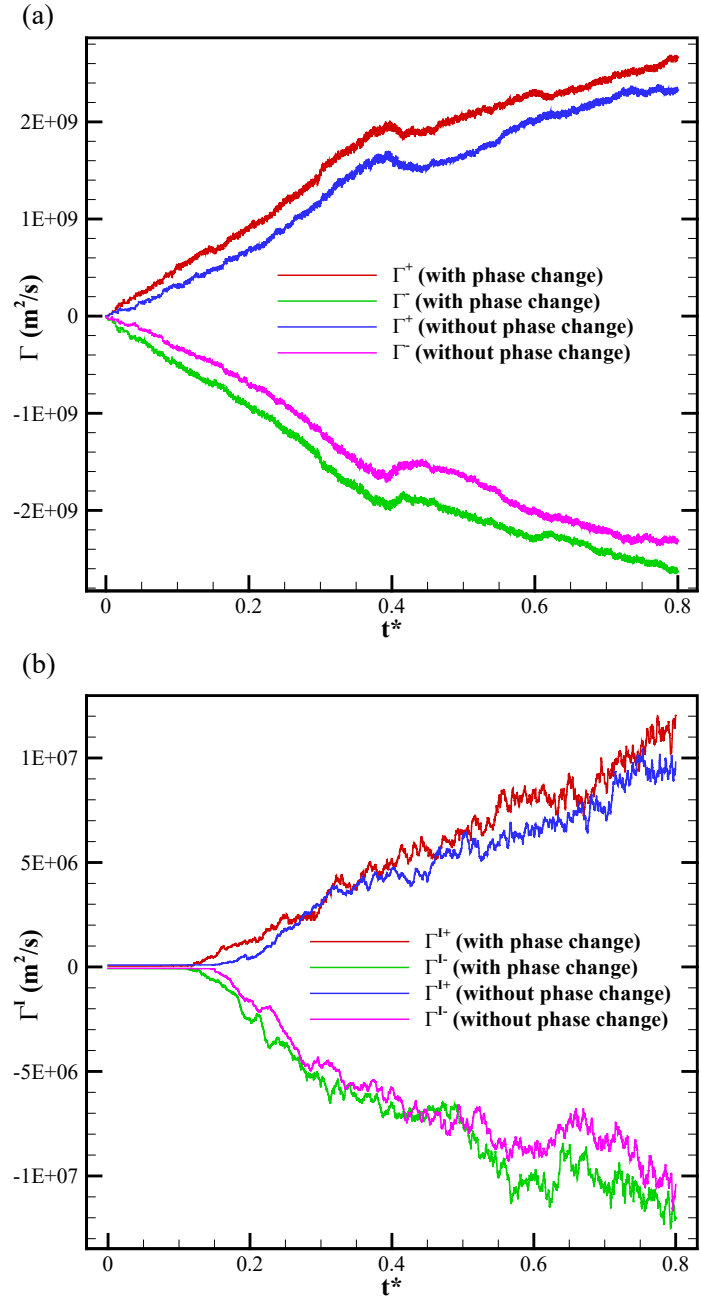

Figure 11. Effect of phase change on the circulation deposition for the entire domain (a) and droplet interface (b).

The influence of the phase change on the flow field around the droplet eventually leads to the difference in the evolution of the droplet interface. To illustrate more clearly, Figure 12 shows the comparison of the zero level-set contours for droplets with and without phase change. Droplet breakup dynamics follows the same SIE breakup mechanism in both cases, which is characterized by the formation of KH-based surface waves on the windward side of the droplet [20]. As is shown in Figure 12(a), the number and amplitude of KHI waves formed on the windward droplet surface increase with consideration of phase change. Subsequently, the external n-dodecane vapor flow entrainment causes the KHI waves to move towards the downstream direction along the windward surface and the liquid carrying KHI waves on the surface of the droplet get accumulated and formed as the thin sheet [21], which can be seen in Figure 12(b). However, as opposed to the counterpart without phase change, the increase of wavenumber and amplitude causes the thin sheet formed earlier and farther from the droplet equator.

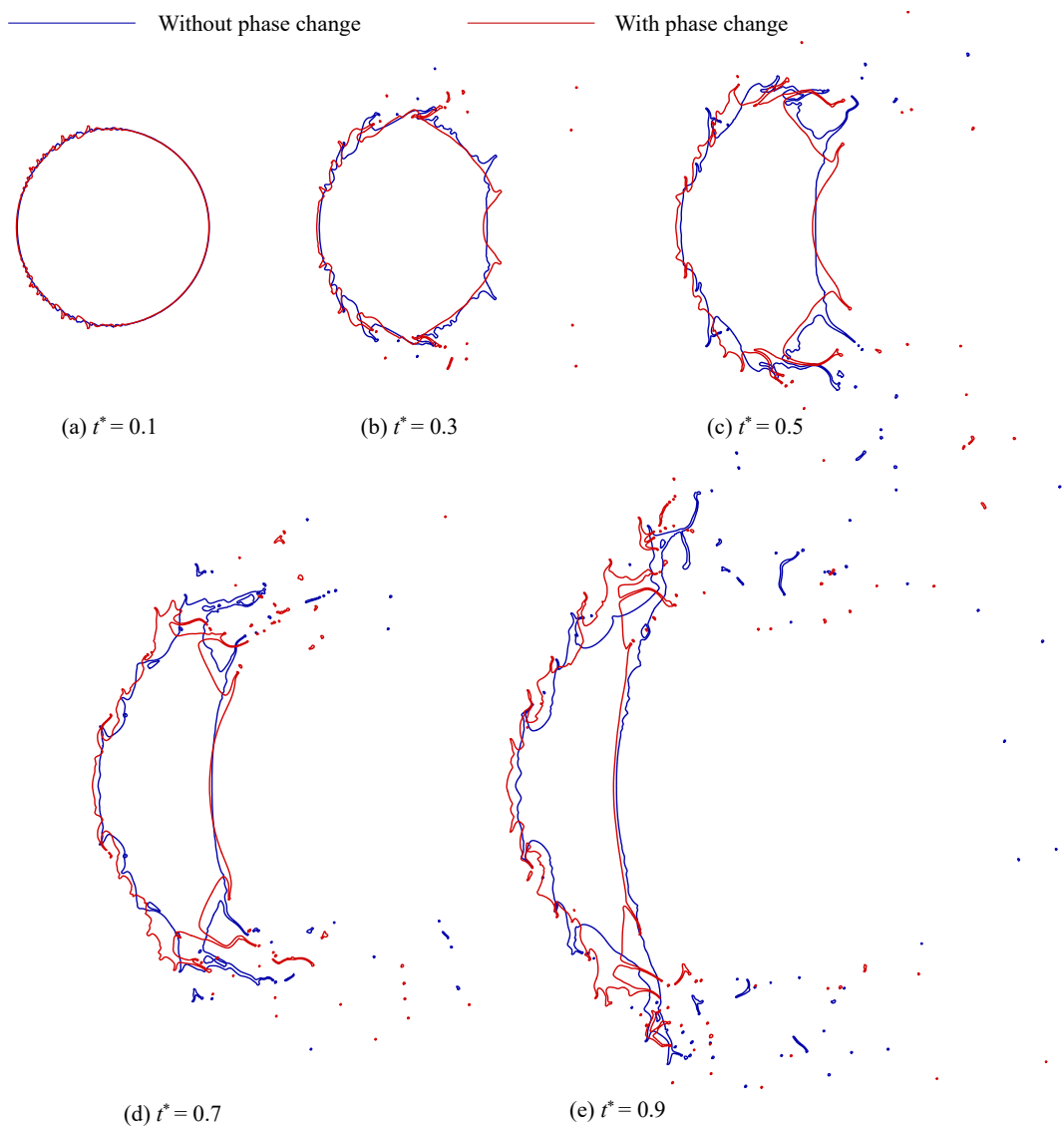

(a) $t^* = 0.1$  (b) $t^* = 0.3$  (c) $t^* = 0.5$

(d) $t^* = 0.7$  (e) $t^* = 0.9$

Figure 12. Comparison of the droplet morphology evolutions with and without phase change.

It is worth noting that the perturbations caused by KHI waves disappear on the leeward side of the droplet with consideration of phase change, which indicates that the KHI waves on the leeward droplet surface are suppressed. According to the previous discussion, the vaporization of droplet impacted by a shock wave occurs mainly on the leeward side of the droplet. Therefore, the suppression of the KHI waves on the leeward droplet surface is mainly caused by vaporization. In addition, the droplet with phase change undergoes deformation due to flattening which causes the internal flow directed towards the equator region leading to the sheet formation augments closer to the front stagnation point, which has a significant effect on the centroid position of the droplet.

Figure 13 quantitatively shows the evolutions of the dimensionless cross-stream diameter and streamwise diameter with and without consideration of the phase change, respectively. On the one hand, the dimensionless streamwise diameter exhibits a continuous decrease due to the flattening of the droplet, which is mainly caused by the pressure difference between the windward and leeward sides of the droplet. On the other hand, due to the formation and stretching of the ligaments at the equator of the droplet, the dimensionless cross-stream diameter shows a continuous increasing trend, and the growth rate is also increasing. However, although qualitatively similar, the two cases are quite different quantitatively. With the influence of phase change, there is a minor increase on the dimensionless streamwise diameter, which suggests that the flattening of the droplet is weakened. Simultaneously, a significantly reduction is found on the dimensionless cross-stream diameter, which shows the stretching of the ligaments becomes also weaker than the case without phase change.

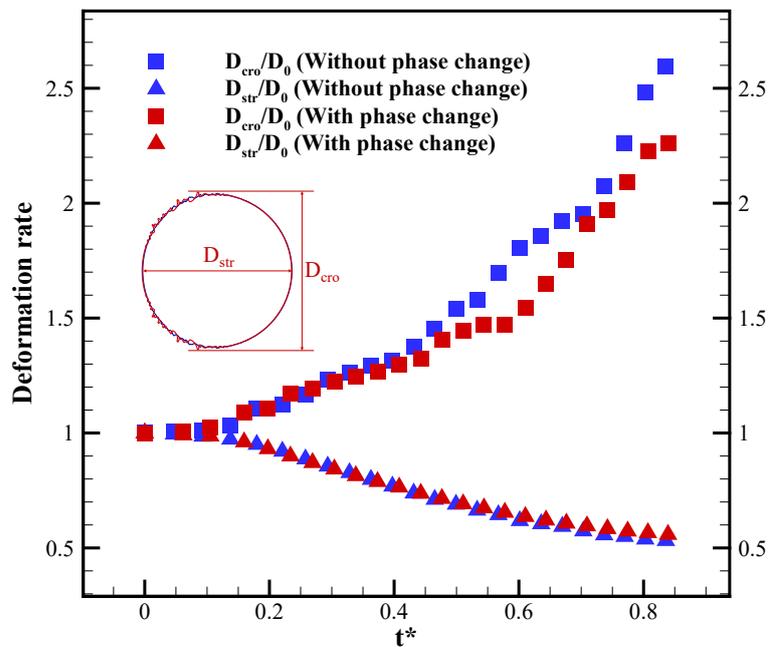

Figure 13. The evolution of the dimensionless cross-stream diameter ($D_{cro}/D_0$) and streamwise diameter ($D_{str}/D_0$) with and without phase change.

In order to quantitatively study the effect of phase change for n-dodecane droplets under the impact of shock waves, Figure 14 plots the droplet's streamwise center-of-mass drift, velocity, acceleration, and drag coefficient evolutions with and without phase change, respectively. Following Meng and Colonius [22], the expression for streamwise dimensionless center-of-mass velocity and acceleration can be derived as

$$\begin{cases} u_c = \dfrac{\int_\Omega u\alpha_l \rho_l dV}{\int_\Omega \alpha_l \rho_l dV}, \ u^* = \dfrac{u_c}{u_g} \\ a_c = \dfrac{\dfrac{d}{dt}\int_\Omega u\alpha_l \rho_l dV}{\int_\Omega \alpha_l \rho_l dV}, \ a^* = \dfrac{a_c D_0}{u_g^2} \end{cases}, \qquad (29)$$

and the drag coefficient is defined as

$$C_d = \dfrac{2F_d}{\rho u^2 S} = \dfrac{2m a_c}{\rho_g (u_g - u_c)^2 D_0}, \qquad (30)$$

where $\alpha_l \rho_l$ denotes the liquid partial density, $\rho_g$ and $u_g$ are the post-shock gas density and velocity, $m$ and $D_0$ are the droplet mass and initial diameter, respectively. In both cases, the law of accelerated motion of the droplet propulsion process is consistent. Droplets are not propelled downstream with a constant acceleration, which will increase as the flattening of the droplet occurs. In the early stages ($t^* < 0.5$), the cross-stream diameter of the droplet remains almost constant, the droplet can be regarded as a rigid body moving with a constant acceleration. However, with the flattening of the droplet and the growth of the liquid sheet, the raise of cross-stream diameter leads to the increase of the windward area, which in turn enhances the aerodynamic force of the droplet.

As is shown in Figure 14, the obvious difference between two cases is the addition of phase change results in a reduction of center-of-mass drift, velocity, acceleration, and drag coefficient, it is not hard to understand. In the first place, the droplet with phase change leads to the sheet formation augments closer to the front stagnation point, which results in a decrease in the location of the droplet's centroid. In addition, the drag coefficient is mainly determined by the aerodynamic force on the droplet, which depends on the size of the windward area. Thus, according to the previous analysis, phase change leads to a weakening of the stretch of the ligaments at the droplet's equator and a significantly reduction on the dimensionless cross-stream diameter, which cause the decrease of the size of the windward area. The reduction of the center of mass and drag coefficient of the droplet together lead to the decrease of the streamwise dimensionless velocity and acceleration.

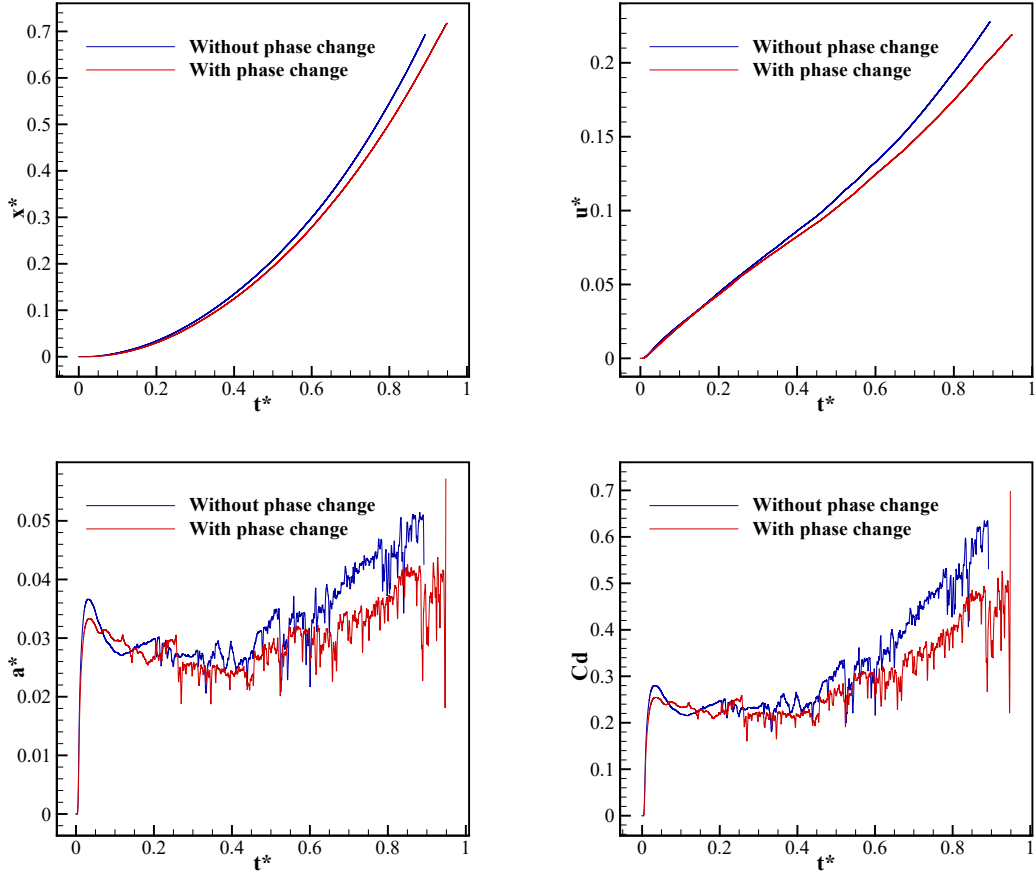

Figure 14. Comparison of the streamwise dimensionless center-of-mass drift, velocity, acceleration, and drag coefficient evolutions with and without phase change.

5.2. Effect of Mach number on shock-droplet interaction with phase change

In the following sections, we compare the cases with various shock Mach number with respect to the droplet morphology evolutions, phase change, and statistical laws of center-of-mass motion. In this paper, the deformation dynamics of droplets with three various of Mach numbers are considered. All three cases consider the effect of phase change, the rest of the conditions remain the same, and the initial conditions for post-shock n-dodecane vapor at three various of Mach numbers calculated by Rankine-Hugoniot relation [56] are shown in Table 4.

Table 4. The initial conditions for post-shock n-dodecane vapor at various of Mach numbers.

| Mach number | $\rho$(kg/m$^3$) | $p$(Pa) | $u$(m/s) | $T$(K) |
| --- | --- | --- | --- | --- |
| 1.37 | 8.66 | $1.89\times 10^5$ | 100.01 | 508.21 |
| 1.47 | 10.15 | $2.35\times 10^5$ | 123.45 | 510.49 |
| 1.57 | 11.81 | 2.48×10$^5$ | 145.94 | 512.85 |

Figure 15 plots the droplet interface evolution for three various of Mach numbers, three cases have the same breakup mechanism because all cases belong to the range of low Ohnesorge number and high Weber number. On the dimensionless time scale, the early-stage droplet deformation for various Mach number cases is compared in Figure 15(a). Obviously, the KHI waves on the windward side of the droplet are enhanced as the Mach number increases, which is manifested in the increase of the wavenumber. In addition, the flattening of the droplet becomes weaker as Mach number increases, which is manifested by an increase in the deformed droplet streamwise diameter. As is shown in Figure 16, the dimensionless droplet streamwise diameter increases significantly in the late stages of droplet breakup. Furthermore, another distinct characteristic of droplet breakup at different Mach numbers is the reduced thickness at the edge of the droplet. In our opinion, this is most likely due to the enhanced shear and stretching at the equator of the droplet as the Mach number increases. In general, as the Mach number increases, the shear stripping of the liquid sheet plays a more dominant role in the deformation and breakup process than the flattening of the droplet under the SIE breakup mechanism.

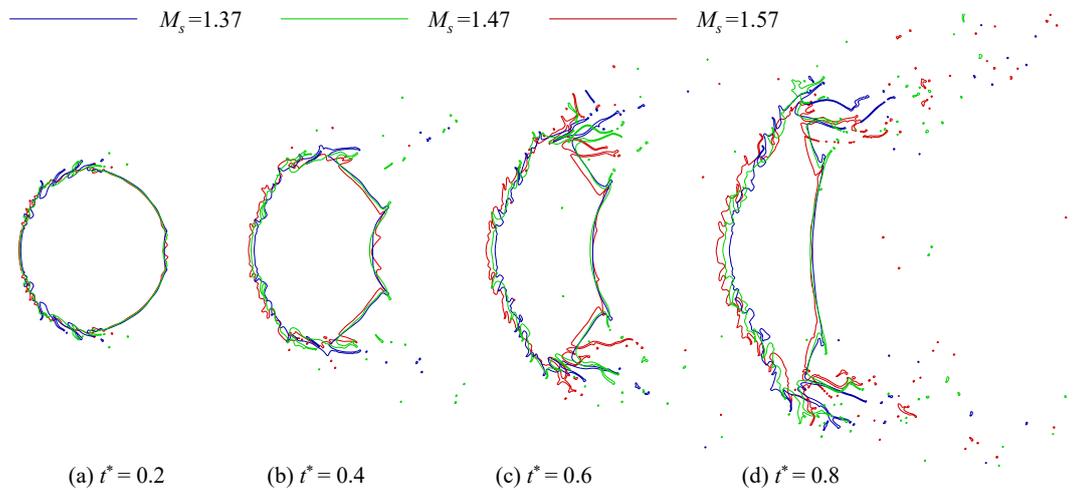

(a) $t^* = 0.2$  (b) $t^* = 0.4$  (c) $t^* = 0.6$  (d) $t^* = 0.8$

Figure 15. Comparison of the droplet morphology evolution with three various of Mach numbers.

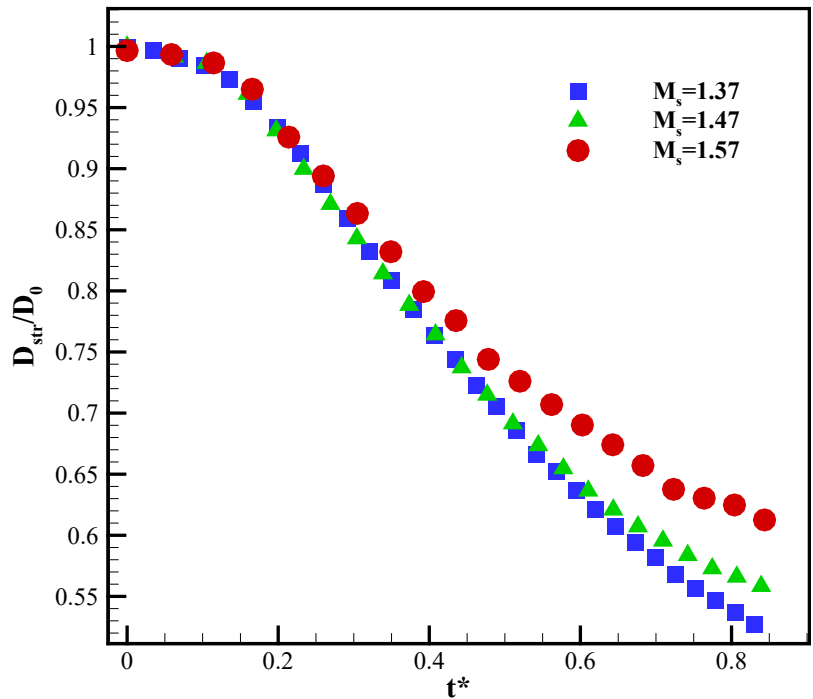

Figure 16. Comparison of the evolution of the dimensionless streamwise diameter with three various of Mach numbers.

Next, let us focus on the phase change of droplets at various of shock Mach numbers. Figure 17 shows the comparison of mass flux distribution of droplet interface with three various of shock Mach numbers at the same dimensionless time. As the shock Mach number increases, the liquefaction on the windward side is significantly enhanced. In addition, the range of liquefaction on the surface of the droplet is gradually expanding from the windward side to the leeward side. As we discussed earlier, the reflection of the incident wave on the windward side forms a high temperature and pressure region where n-dodecane vapor undergoes liquefies on the windward side because the pressure in this region is greater than the saturated vapor pressure at the corresponding temperature. As the shock strength increases, the pressure and temperature on the windward side of the droplet also increases. However, the increase in pressure is much higher than the increase in temperature, which causes the liquefaction rate on the windward side of the droplet is increased. Figure 18 plots the droplet dimensionless mass evolutions with various shock Mach numbers, the overall vaporization rate of the droplet decreases with the increase of the shock Mach number, such that the droplet exhibits liquefaction at higher Mach numbers. Of course, if we increase the initial temperature of the vapor, the droplets will still show vaporization in the overall.

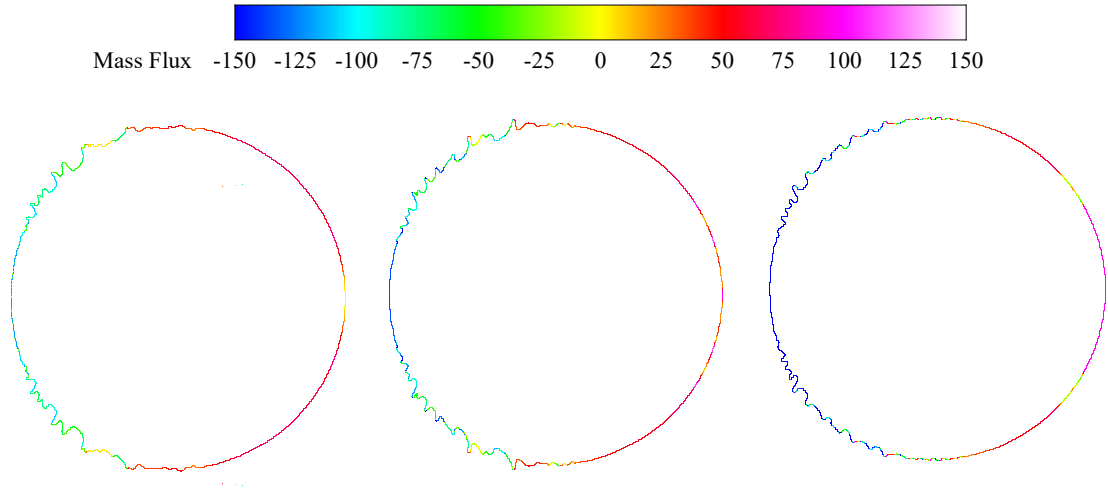

Figure 17. Mass flux distribution along the deformed droplet interface with various Mach number at $t^* = 0.1$.

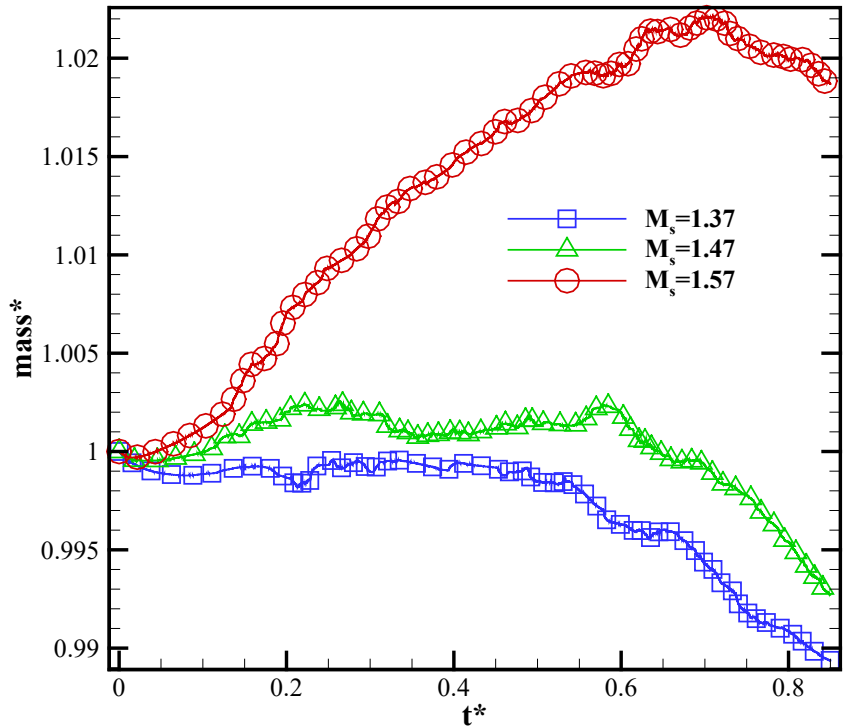

Figure 18. Comparison of the droplet dimensionless mass evolutions with three various of Mach numbers.

Furthermore, Figure 19 compares the phase change droplet center-of-mass drift, velocity, acceleration, and drag coefficient evolutions with three various of Mach numbers. In a higher Mach number, the droplet drift has an obvious tendency to be smaller, which is caused by the droplet morphology difference. As is shown in Figure 15, the streamwise diameter of the droplet and the position of the front stagnation point are more forward, which has a significant impact on the centroid position. The dimensionless center-of-mass velocity and acceleration increase with Mach number, especially in the late stages, which is agree with the results of Meng and Colonius [22] in the cases of droplet breakup without phase change. Moreover, the drag coefficient remains essentially the same at different Mach numbers, which indicates that the wave drag also does not significantly alter the drag coefficient with the consideration of phase change.

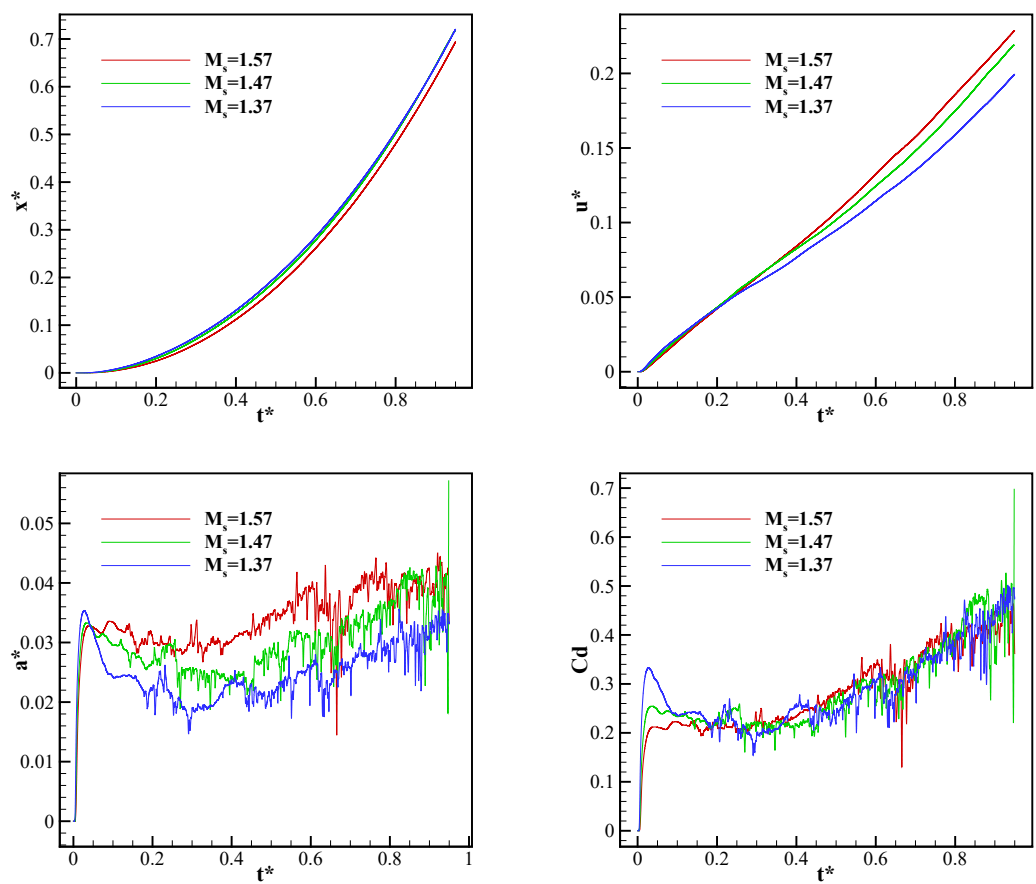

Figure 19. Comparison of the streamwise dimensionless center-of-mass drift, velocity, acceleration, and drag coefficient evolutions with three various of Mach numbers.

## 6. Conclusion

In this paper, a numerical investigation is carried out on the interaction of shock wave and n-dodecane droplet with phase change. A fully conservative sharp-interface method for compressible multiphase flows with phase change is employed as the numerical method. For n-dodecane fluids, the equations are closed using a real-fluid EOS based on Helmholtz-energy to obtain more accurate thermodynamic variables physically. Firstly, a water droplet impacted by an incident shock without phase change is introduced to test the validation of the numerical method and grid convergence, the results show good agreement with the experiments. Next, a comparative study on the shock wave and n-dodecane droplet interaction with and without phase change model is conducted to investigate the effects of phase change for droplet propulsion, deformation, and fragmentation. It is found that the impact of the shock wave weakened the vaporization and caused the liquefaction of the windward side of the droplet. With the effect of phase change, more vorties are generated, the KHI waves on the windward surface are enhanced, but inhibited by vaporization on the leeward side. Furthermore, both the flattening and shearing of the droplet are suppressed, which cause the reduction of center-of-mass drift, velocity, acceleration, and drag coefficient. Lastly, we perform a high-resolution numerical simulation of the deformation, phase change and propulsion of n-dodecane droplets with different shock Mach numbers. The results show that shear stripping plays a more dominant role in droplet breakup than flattening under the SIE breakup mechanism with the increase of shock Mach number. The increase in shock strength significantly enhances the liquefaction effect on the windward side of the droplet. Furthermore, the shock strength does not significantly alter the drag coefficient with the consideration of phase change.